\documentclass[12pt,a4paper,onecolumn]{article}

\usepackage[T1]{fontenc}
\usepackage[utf8]{inputenc}
\usepackage{graphicx}
\usepackage{amsmath}
\usepackage[usenames]{color}
\usepackage{cite}
\usepackage{authblk}
\DeclareUnicodeCharacter{200B}{\empty}

\title{Exciton screening in C$_{60}$ and PTCDA complexes. TDDFT calculations with GGA and hybrid functionals.}
\author[1]{N.L. Matsko}
\author[1,2,3]{Mahmoud A. Salem}
\affil[1]{\footnotesize Bogoliubov Laboratory of Theoretical Physics, Joint Institute for Nuclear Research, 141980 Dubna, Russia}
\affil[2]{\footnotesize Department of Physics, Faculty of Science, Zagazig University, 44519 Zagazig, Egypt}
\affil[3]{\footnotesize The Academy of Scientific Research and Technology, 4262104 Cairo, Egypt}

\begin{document}

\maketitle

\begin{abstract}

Photoabsorption in the low-energy region for C$_{60}$ and PTCDA molecular complexes is studied within linear response TDDFT. For the PBE, B3LYP and HSE exchange-correlation (xc) functionals the dependence of the accuracy of the exciton energy on the electron-hole separation is analyzed. Particular attention is paid to the charge-transfer (CT) excitons. The inclusion of non-local exchange using hybrid functionals increases the accuracy of calculations for short-range excitons, however, the accuracy of hybrid functionals decreases significantly for long-range excitons. Moreover, as the exciton radius approaches the "screening length"\ , the simpler PBE functional gives more accurate excitonic energies than the mentioned hybrid functionals.
\end{abstract}

\section{Introduction}

The linear response TDDFT methods have become a powerful tool for calculating excitations of a many-electron system. Solving the Kohn-Sham equations for the ground state with a Hamiltonian $H_{scf}$ we find eigenvalues $\epsilon_n$ and eigenvectors $|\psi_n>$. Applying first-order perturbation theory we can write an analogue of the Kohn-Sham equations in the presence of an external disturbing potential \cite{baroni}:

\begin{equation} \label{1} (H_{scf}-\epsilon_n)|\Delta\psi_n>=-(\Delta V_{scf}-\Delta\epsilon_n)|\psi_n>
\end{equation}

where $\Delta V_{scf}(r)$ - is the first-order variation of the Kohn-Sham effective potential:

\begin{equation} \label{2} \Delta V_{scf}(r)= \Delta V_{ext}(r)+e^2\int\frac{\Delta\rho(r')}{|r-r'|}dr'+\frac{\partial V_{xc}}{\partial\rho}\Delta\rho(r)
\end{equation}

\begin{equation} \label{3} \Delta\rho(r)=2\sum_n^{occup}\sum_{m\ne n}\psi_n^*(r)\psi_m(r)\frac{<\psi_m|\Delta V_{scf}|\psi_n>}{\epsilon_n-\epsilon_m}
\end{equation}

The $\Delta\rho(r)$ in the (\ref{3}) depends on $\Delta V_{scf}(r)$ and therefore it depends on the chosen $V_{xc}$. In turn, $\Delta V_{scf}(r)$ in the (\ref{2}) depends on the $\Delta\rho(r)$. Equations (\ref{1})-(\ref{3}) are solved self-consistently. The solution depends on the xc functional determining the interaction between changes in electron density $\Delta\rho(r)$ in (\ref{1}). In general, the interaction decreases as $\epsilon^{-1}(r,\omega)/r$, where the response $\epsilon$ depends on distance and frequency. For small $r$, the main non-local contribution to the $\epsilon(r,\omega)$ is usually determined by the exchange interaction, which eliminates DFT self-interaction error. As $r$ increases and more particles are involved, the contribution of collective screening effects increases and becomes comparable to the exchange one. In a conductive system, at a distance of the screening length, the static charge is completely screened. In a semiconductor, screening is caused by valence electrons with lower mobility, which generally do not completely screen the charge in the system, but weaken it. In both cases, on the scale of the screening length, the correlation contribution becomes comparable to the exchange one. The screening length acts as an important parameter, distinguishing between short-range and long-range order in the system. For the former the system is well described within the framework of the local one-electron approximation. And for the long-range limit the system can be described within the framework of collective variables (like in the RPA \cite{pines}).

The simplest way is to select the xc contribution corresponding to a homogeneous electron gas of the given electron density. For example, it could be GGA type functional such as PBE \cite{pbe1,pbe2}. In this case, the accuracy of Frenkel exciton calculations is often around 0.5-0.7 eV \cite{maier}. A standard method for more accurate accounting of nonlocal effects is to use hybrid xc functionals containing a direct part of the Fock exchange. In turn, the exchange interaction could be range-separated. Typically, hybrids allow to increase the accuracy of Frenkel exciton calculation to 0.1-0.4 eV \cite{maier,medves,petersilka}. Most of the works on TDDFT electron calculations use hybrid-type functionals, however most of them consider relatively small systems \cite{maier,Kozma,Casanova,drew,tawada}. In the case of a more extended systems with delocalized excitons being considered (such as CT and Rydberg excitons), hybrid functionals can give an error of up to 1 eV \cite{drew,tawada,stein,campetella}. Thus, to describe the long-range tail of the electron–hole interaction it is not enough to take into account only the exchange interaction. Adequate treatment of correlations is also required \cite{kresse}. The selected xc functional must correctly capture the system’s response at distances comparable to the characteristic radius of the excitation.

It is also worth noting the works considering excitons in molecular systems with a radius of 10 \AA\ and more \cite{ramanjanappa,kozlowska,karolewski,majeed,marmolejo,SAFA2024e23813}. In such systems, the efficiency of various xc functions looks different. In the works \cite{ramanjanappa,kozlowska} it is concluded, that conventional GGA functionals display high performance for CT excitation energies with an accuracy of about 0.1-0.2 eV. In the work \cite{kozlowska} authors point that the worst performance was observed for the range-separated hybrids. The above can be explained as follows. In GGA functionals the balance of exchange and correlations at large distances is more accurate than in hybrid functionals. GGA inherently satisfies the asymptotics for the xc contribution \cite{pbe1,pbe2}. In particular, exchange and correlations in GGA (and LDA) diverge on large scales, but with opposite signs and compensate each other. In hybrid functionals, the GGA exchange is partly replaced by the exact Fock exchange, thus the balance of the exchange and correlation parts on a large scale is disrupted accordingly.

In this work, we calculate the electronic excitations in complexes of C$_{60}$ and PTCDA in the frame of linear response TDDFT. For the given systems it is of interest to study the role of screening in describing of long-range excitons, as well as the efficiency of various xc functionals. First of all, we can make a rough estimate of the characteristic distance, "screening length" for a system consisting of C$_{60}$ molecules. This should correspond to the screening length in graphene across the layers that equals to 1.2 nm \cite{miyazaki},\cite{hachen}. Thus the particular attention is paid to the calculations of the long-range charge-transfer excitons with comparable size. Calculations are carried out using PBE, B3LYP and HSE functionals.

\section{Computational Details}

DFT calculations with the plane-wave basis were performed in Quantum Espresso package \cite{Giannozzi_2009,Giannozzi_2017} and the optimized Norm-Conserving Vanderbilt (ONCV) Pseudopotentials \cite{PhysRevB.88.085117} were implemented. The plane-wave cutoff energy was set to 35 Ry. Computations were performed for the supercells with the vacuum region of 10 \AA\ between replicas from neighboring supercells. The GGA exchange-correlation functional with Perdew-Burke-Ernzerhof (PBE) parametrization \cite {pbe1}, Becke’s three-parameter hybrid exchange B3LYP functional \cite{stephens1994ab}, the range-separated hybrid HSE \cite{10.1063/1.2204597}. For the linear-response TDDFT calculations, the Lanczos method implemented in Quantum Espresso \cite{PhysRevResearch.5.023089,malcioglu,GE20142080} was used to calculate the absorption spectra.

The geometries of molecular structures under study were relaxed until atomic forces became less than $10^{-5}$ Ry/atom. The calculated equilibrium distance between C$_{60}$ molecules (distance between the nearest C atoms belonging to different C$_{60}$ molecules) equals to 4 \AA\ for PBE functional, 3.8 \AA\ for PBE with vdW D3 correction \cite{grimme} and 3.3 \AA\ with vdW TS correction \cite{tkachenko}. The experimental value for the solid C$_{60}$ is indicated as 2.5-3 \AA\ \cite{roy,dressel,reddy}. Equilibrium distance variations for different functionals and difference from the experimental value is apparently explained by problems with the correct account for dispersion forces. The dynamically screened vdW interaction contains a significant surface plasmon contribution in the case of C$_{60}$ \cite{tao,klimov,schmeits}. Simple DFT vdW corrections are not accurate enough in this case. Photoabsorption spectra were calculated for the single C$_{60}$ molecule, two, three and four C$_{60}$ molecules arranged in a line with intermolecular distances of 3.8, 3.3, 2.8 \AA.

PTCDA is a planar molecule with a central perylene core and four carboxylic dianhydride groups. The main types of packing of PTCDA molecules in a monolayer are $\alpha$ and $\beta$ polymorph structures \cite{ogawa,han}. Neighboring molecules in a monolayer are rotated approximately 90 degrees relative to each other. Molecules belonging to different monolayers form stacks tilted at an angle of 10–12 degrees to the monolayer plane in the case of an ideal crystal \cite{forrest,shen}. Interlayer (stacking) distance between molecules equals to 3.2 \AA\ \cite{shen}. Calculations show that the excitations inside the monolayer are noticeably weaker in intensity compared to the ones in the stack \cite{shen,zhang,mazur}. Therefore, we will consider only excitations in the stack. Monolayers in a real system can shift relative to each other due to stress, defects, edge effects \cite{cochrane}. Therefore, we considered the two characteristic geometries of the stack. In the first case, the stack contains molecules positioned one above the other and rotated relative to each other by 90 degrees (inset in Fig. 2b). In the second case, the stack contains molecules positioned one above the other without rotation (inset in Fig. 2c). The specified stack geometries containing 2 and 3 PTCDA molecules were considered for the photoabsorption calculations. The equilibrium distance between PTCDA molecules for the calculations with PBE functional is 3.35 \AA\ for the first stack and 3.4 \AA\ for the second stack.

In order to verify the accuracy of the model we made test calculations of the main low energy photoabsorption peaks for three simple systems. The calculated peak in the CO molecule is 8.1 eV, 8.39 eV, 8.46 eV and 8.91 eV for the PBE, B3LYP, HSE and Hartree-Fock (H-F) functionals respectively. Experimental value is 8.73 eV \cite{tanaka}. The two main peaks in the CH$_4$ molecule are 8.93 and 9.5 eV, 9.52 and 10.33 eV, 9.75 and 10.63 eV, 10.84 and 12.99 eV for the PBE, B3LYP, HSE and H-F functionals. While the experimental values are 9.5-9.76 and 10.3-10.33 eV \cite{lee1983,chen2004}. The main peak in the C$_6$H$_6$ molecule is 6.27 eV, 6.52 eV, 6.75 eV, 6.85 eV for the PBE, B3LYP, HSE and H-F functionals. Experimental values is 6.8-6.94 eV \cite{inagaki,koch}. One can see the severe underestimation of the excitation energy for PBE and noticeably less underestimation for hybrids functionals. Moreover, for the CO and C$_6$H$_6$ molecules the H-F functional gives the best value. For more extended systems the H-F functional begins to significantly overestimate the excitation energy.
These results are consistent with other calculations \cite{yabana,walker,qian,hanasaki}, demonstrating a significant improvement in accuracy of photoabsorption energy for hybrid functionals in simple systems.

\section{Results}

{\bf C$_{60}$ complexes}

Absorption-edge and the low-energy photoabsorption peaks in the solid C$_{60}$ have the following structure. The weak diffuse peaks corresponding to Frenklel excitons in a single C$_{60}$ molecule are observed at the energy of 1.8-2.1 eV \cite{knupfer,hartmann,reber,salim}. These peaks are essentially made of states of a Frenkel excitonic nature, strongly coupled to vibrational modes \cite{knupfer,hartmann,lof,guss,kobayashi}. The corresponding excitations are dipole-forbidden and can be optically observed due to interactions with atomic oscillations. Accordingly, such diffuse peaks are not reproduced in our calculations. Higher in energy there is a peak in the region of 2.7-2.8 eV, as well as weak signals are in the region of 2.2-2.5 eV \cite{salim,chilu,sharma,skumanich}. These excitations are absent in the case of a single C$_{60}$ molecule and are not localized on a single molecule. Most often, these excitations are referred to as CT excitons, when the electron is located on one C$_{60}$ and the hole is located on the other one, less often to Wannier excitons \cite{muhammed} and to mixed CT and Frenkel type excitons \cite{knupfer, kobayashi}. These excitations in the region of 2-3 eV are most interesting for us. Higher in energy Frenkel excitons at 3.8 and 4.9 eV are observed \cite{yasumatsu}.

\begin{figure}[h]
\centering
\includegraphics[width=0.7\textwidth]{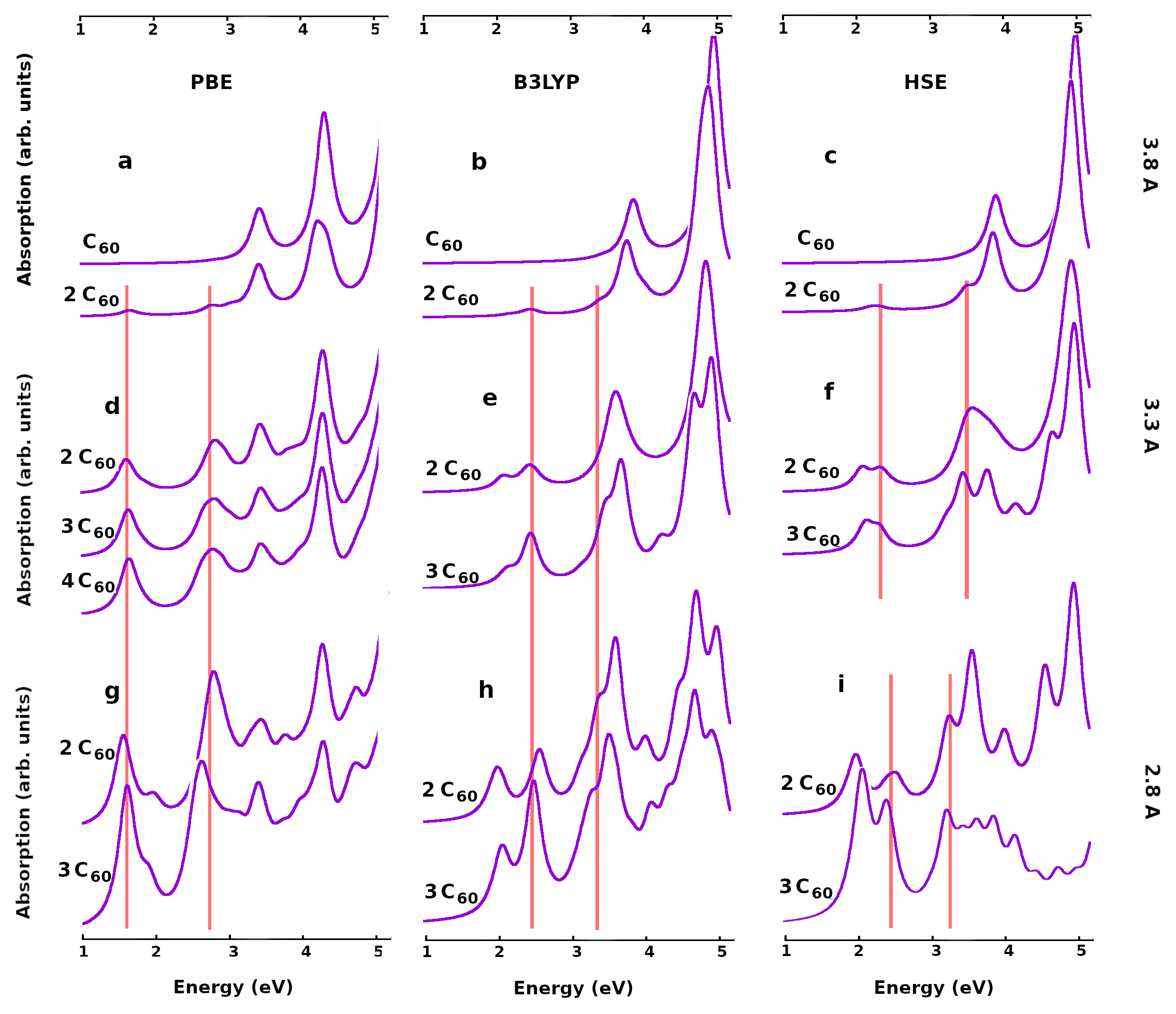}
\caption{C$_{60}$ photoabsorption calculations for PBE, B3LYP and HSE functionals. $\bf{a}$, $\bf{b}$, $\bf{c}$ - calculations of one C$_{60}$ molecule and two C$_{60}$ molecules with an intermolecular distance of 3.8 \AA. $\bf{d}$, $\bf{e}$, $\bf{f}$ - calculations of two and three C$_{60}$ molecules with an intermolecular distance of 3.3 \AA. The absorption of four C$_{60}$ molecules for PBE functional is also given. $\bf{g}$, $\bf{h}$, $\bf{i}$ - calculations of two and three C$_{60}$ molecules with an intermolecular distance of 2.8 \AA. The vertical lines indicate the discussed collective exciton peaks.}
\label{Fig1}
\end{figure}

The calculated photoabsorption spectra are presented in Fig. 1. The curves are organized into three columns, corresponding to calculations with the PBE, B3LYP and HSE functionals. The curves in Fig. 1 are also organized into three rows. The first row shows calculations for a single C$_{60}$ molecule and for two C$_{60}$ molecules with an intermolecular distance of 3.8 \AA. The second and the third row correspond to calculations for 2 and 3 C$_{60}$ with an intermolecular distance of 3.3 \AA\ and 2.8 \AA. The photoabsorption curves for a single C$_{60}$ molecule show two prominent peaks at 3.45 and 4.35 eV for PBE, 3.88 and 4.98 eV for B3LYP, 3.92 and 5.02 eV for HSE. The experimental values ​​are 3.8 and 4.9 eV. PBE exhibits a typical underestimation of energy values, while hybrid functionals yield a more accurate result with a slight overestimation.

The two new peaks appear in the calculation when we consider more than one C$_{60}$ molecule. These low-energy peaks emerge only when the external perturbation is polarized along the axis connecting fullerene molecules. Thus, the corresponding excitations have dipole components directed along the C$_{60}$-C$_{60}$ axis. We will call these peaks collective since they are related to excitations of more than one C$_{60}$. The experimental energies for these collective excitons are 2.2-2.5 and 2.7-2.8 eV. They are usually associated with Frenkel and CT excitons. In the case of two C$_{60}$ molecules with a distance of 3.8 \AA, the values ​​of the peaks are 1.66 and 2.81 eV for PBE, 2.46 and 3.5 eV for B3LYP, 2.28 and 3.58 eV for HSE. As can be seen from Fig. 1 $\bf{a}$, $\bf{b}$, $\bf{c}$ the indicated peaks have a weak intensity.

When the intermolecular distance decreases to 3.3 \AA\ (Fig. 1 $\bf{d}$, $\bf{e}$, $\bf{f}$), the intensity of the two collective peaks increases noticeably, indicating an increase in intermolecular charge transfer. The single molecule peaks change rather slightly. A further effect observed as the molecules approach is the splitting of the lowest-energy peak, which is more pronounced for the hybrid functionals.
The calculated energies of the collective peaks are 1.61 and 2.84, 1.63 and 2.82, 1.64 and 2.78 eV for PBE in case of two, three and for C$_{60}$ molecules respectively. 2.46 and 3.45, 2.47 and 3.52 eV for B3LYP in case of two and three C$_{60}$ molecules. 2.1-2.32 and 3.3, 2.14-2.25 and 3.2 eV for HSE in case of two and three molecules.

With a further decrease in the intermolecular distance to 2.8 \AA\ (Fig. 1 $\bf{g}$, $\bf{h}$, $\bf{i}$) the intensity of the collective peaks increases even more. The splitting of the lower peak also increases giving two distinct peaks for B3LYP and HSE. The energies are 1.55 and 2.8, 1.6 and 2.65 eV for PBE in case of two and three C$_{60}$ molecules respectively. 2, 2.58 and 3.2 eV, 2.1, 2.5 and 3.3 eV for B3LYP in case of two and three C$_{60}$ molecules. 2, 2.52 and 3.28 eV; 2.07, 2.4 and 3.24 eV for HSE in case of two and three molecules. In addition, incomplete convergence for energies above 3-4 eV can be noted for the photoabsorption curve of three C$_{60}$ molecules in Fig. 1$\bf{i}$.

Calculations show, that PBE gives a clear peak in the region of 2.65-2.8 eV corresponding to the experimental value. The B3LYP and HSE functionals overestimate the value by about 0.5-0.7 eV, showing a peak in the region of 3.2–3.4 eV. For peaks in the range of 2.2–2.5 eV, the situation is opposite. The B3LYP and HSE functionals provide fairly accurate values, while the PBE gives a strong underestimation. This change is apparently explained by the different nature of the excitations. In most studies \cite{kobayashi,shirley,munn}, the peak in the region of 2.7–2.8 eV is associated with a delocalized exciton, usually of a CT type, while the lower excitations are usually associated with localized excitons of Frenkel type. The size of the CT exciton should be comparable to or greater than the distance between the centers of two C$_{60}$ molecule. This distance is 10-11 \AA, which corresponds to the "screening length" estimate in the introduction.

{\bf PTCDA complexes}

\begin{figure}[h]
\centering
\includegraphics[width=0.6\textwidth]{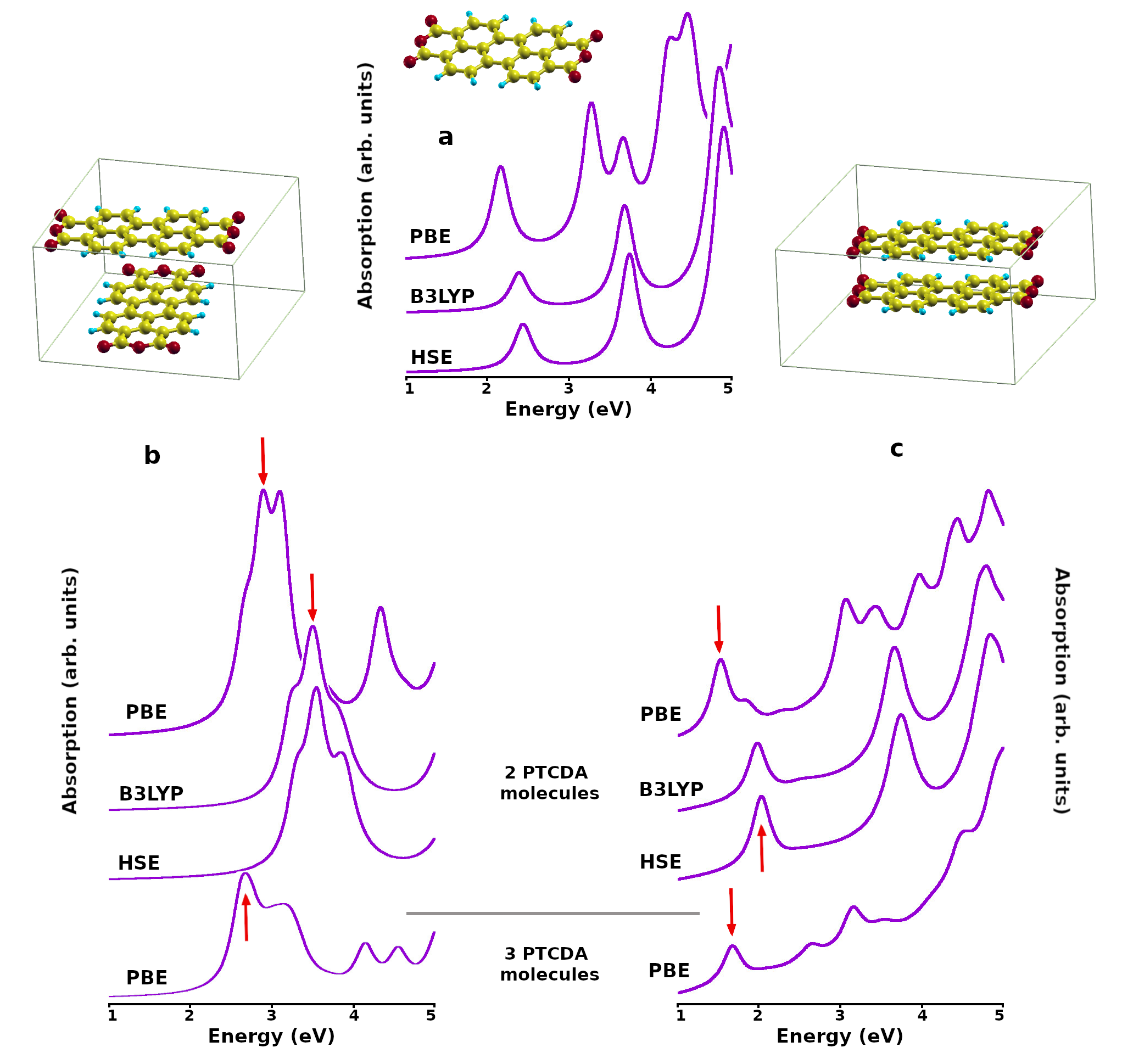}
\caption{PTCDA photoabsorption calculations with PBE, B3LYP and HSE functionals. $\bf{a}$ - PTCDA single molecule calculations. $\bf{b}$ - calculations for 2 and 3 PTCDA molecules in the stack, when the upper molecule is rotated 90 degrees relative to the lower one. $\bf{c}$ - calculations for 2 and 3 PTCDA molecules in the stack, when molecules are not rotated.
Since the perturbing field for the systems in graphs $\bf{b}$ and $\bf{c}$ is directed perpendicular to the plane of the PTCDA molecule, the Frenkel exciton in the region of 2.2 eV is not excited.}
\label{Fig2}
\end{figure}

Fig. \ref{Fig2} $\bf{a}$ shows the photoabsorption calculations in PTCDA single molecule with PBE, B3LYP and HSE functionals. The main low-energy peak of the Frenkel exciton has values 2.16, 2.38 and 2.41 eV respectively.
In the PTCDA crystall this peak has a close frequency because of the weak vDW intermolecular interaction and demonstrates an experimental value of 2.2-2.25 eV \cite{shen,mazur,gangilenka}. So the B3LYP and HSE functionals overestimate the value by approximately 0.2 eV, while PBE leads to an underestimation of the energy by 0.05-0.1 eV. In the calculation, this excitation is observed when the external disturbing field has the same direction as the long axis of the molecule (with a length of 11.2 \AA). Thus, the exciton has a charge distribution along the molecule long axis and should be of comparable size. In the work \cite{shen} the exciton radius is defined as 13$\pm$2 \AA. It can be assumed that a more accurate accounting of the xc contribution in PBE at this scale leads to a more accurate result.

For the stack in Fig. 2 $\bf{b}$ the calculations show an intense low-energy peak at 2.89 eV for PBE, 3.5 eV for B3LYP, and 3.54 eV for HSE when considering two molecules in the stack. The frequency changes to 2.67 eV for PBE when considering three molecules in the stack. The peak is marked with arrows in the figures. The experimental value is 2.5-2.6 eV and the peak is associated with the CT exciton \cite{han,shen,foker}.
For the stack in Fig. 2 $\bf{c}$ the low-energy CT peak has a frequency of 1.5 eV for PBE, 1.97 eV for B3LYP, and 2.02 eV for HSE when considering two molecules in the stack. The frequency changes to 1.66 eV for PBE when considering three molecules in the stack. The peak is marked with arrows in the figures. Photoluminescence experiments show the presence of 4-5 peaks in the region of 1.75-2 eV. These peaks are associated with Frenkel excitons, as well as CT excitons and self-trapped CT excitons \cite{gangilenka,kobitski}. The low-energy peak in Fig. 2 $\bf{c}$ corresponds to CT exciton (non self-trapped) since we do not consider relaxation of the excitated structure. In the work \cite{gangilenka} this CT exciton is attributed to the peak at 1.97-2 eV.

For the stack in Fig. 2 $\bf{b}$ the B3LYP and HSE greatly overestimate the energy of the CT excitation, whereas PBE gives value only 0.1 eV higher. Conversely, for the stack in Fig. 2 $\bf{c}$ B3LYP and HSE give an accurate value, while PBE shows a strong underestimation. In both cases, the CT exciton is considered. The difference is apparently explained by the radius of the excitons. For the stack in Fig. 2 $\bf{b}$, the exciton has a larger radius, the electron and the hole are separated between two molecules, and are also spaced apart in the plane of the molecules themselves. In the paper \cite{shen}, the size of the exciton is defined as 14$\pm$2 \AA. For the stack in Fig. 2 $\bf{c}$ the exciton has a small radius, apparently about the intermolecular distance of 3.2 \AA, with the electron and the hole localized on adjacent molecules one above the other.

\section{Conclusions}

In summary, we conclude that in the systems under consideration a length of 10-15 \AA\ can be interpreted as an analogue of the screening length. For electronic excitations with a comparable radius, the proper consideration of screening effects becomes essential. Taking into account only the correction for non-local exchange, as in simple hybrid functionals, is not sufficient.
For the considered C$_{60}$ and PTCDA molecular complexes the above applies to the low-energy CT excitons, associated with charge transfer between neighboring molecules, as well as to the lowest large-scale intramolecular Frenkel exciton in PTCDA. For these excitations, the B3LYP and HSE functionals overestimate the energy by approximately 0.5 eV. The simpler PBE functional gives an error of about 0.1 eV and is significantly more accurate. This could be explained by the fact that PBE inherently satisfies the asymptotics for the xc contribution averaged over a large scale. In simple hybrid functionals, the PBE exchange is partly replaced by the exact Fock exchange, the sum of the exchange and correlation parts on a large scale changes accordingly, and the corresponding contributions do not compensate each other properly.
The use of more complex exchange-correlation functionals with non-local exchange and correlations should overcome these problems. However, the application of complex xc functionals in large systems poses a significant challenge. Instead, when calculating an exciton with a radius equal to or greater than the "screening length", the PBE functional could be a good alternative. The PBE yields significantly more accurate exciton energies than the simple hybrids, demonstrating accuracy in the region of 0.1 eV.

\section{Acknowledgements}

Calculations were carried out on the Govorun supercomputer of the Joint Institute for Nuclear Research.

\bibliographystyle{unsrt}
\bibliography{reference}

\end{document}